\documentclass[useAMS,usenatbib,usegraphicx]{mn2e}
\usepackage{graphicx}
\usepackage{txfonts}

\title [Stability and secondary resonances in the SR3BP]{Stability and secondary 
resonances in the spatial restricted three-body problem for small mass ratios}
\author[R. Schwarz, \'A. Bazs\'o, B. Funk, and B. \'Erdi]
{R. Schwarz$^{1}$\thanks{E-mail:schwarz@astro.univie.ac.at}, \'A. Bazs\'o$^{1}$,
B. \'Erdi$^{2}$, and B. Funk$^{1}$\\
$^{1}$Institute for Astronomy, University of Vienna, A-1180 Vienna, 
T\"urkenschanzstrasse 17, Austria\\ 
$^{2}$Department of Astronomy, E\"otv\"os University, H-1117 Budapest, 
P\'azm\'any P\'eter s\'et\'any 1/A, Hungary\\
}

\begin{document}

\date{Accepted 1988 December 15. Received 1988 December 14; in original 
form 1988 October 11}

\pagerange{\pageref{firstpage}--\pageref{lastpage}} \pubyear{2002}

\maketitle
\label{firstpage}

\begin{abstract}
This paper is devoted to the study of secondary resonances and the stability of 
the Lagrangian point $L_4$ in the spatial restricted three-body problem for 
moderate mass ratios $\mu$, meaning that $\mu \le 0.0045$. However, we 
concentrated our investigations on small mass ratios $\mu \le 0.001$, which 
represent the mass ratios for stable configurations of tadpole orbits in the 
Solar system.

The stability is investigated by numerical methods, computing stability maps in 
different parameter planes. We started investigating the mass of the secondary; 
from Earth-mass bodies up to Jupiter-mass bodies.
In addition we changed the orbital elements (eccentricity and inclination) of the 
secondary and Trojan body.
For this parameter space we found high order secondary resonances, which are 
present for various inclinations. To determine secondary resonances we
 used Rabe's equation and the frequency analysis.
In addition we investigated the stability in and around these secondary resonances.  

\end{abstract}

\begin{keywords}
celestial mechanics -- minor planets, asteroids -- Solar system: general --
methods: numerical
\end{keywords}

\section{Introduction}
\label{intro}
Trojan asteroids are minor planets librating in the vicinity of the $L_4$ or $L_5$ points 
of a planet; e.g. Jupiter. $L_4$ and $L_5$ are equilibrium points and can be regarded as 
realizations of the Lagrangian triangular solutions of the restricted 
three-body problem which are stable for the mass ratio $\mu \lesssim 0.0385$, 
where $\mu$ is defined as $\mu=m_2/(m_1+m_2)$; $m_1$ and $m_2$ 
are the masses of the primary and secondary bodies.
The first Trojan asteroids were found in a 1:1 mean motion resonance with Jupiter, 
librating around the points $L_4$ or $L_5$ of the Sun-Jupiter system; 
preceding ($L_4$) or following ($L_5$) Jupiter at about $60^{\circ}$.
Today we know approximately 6000 Trojans. Most of them are Jupiter Trojans (5984) 
but also one Earth Trojan, 4 Martian Trojans, one Uranus Trojan and 9 
Neptune Trojans were detected (April 2014).

Several authors have conducted dynamical studies for Trojan asteroids in the 
Solar system \citep[e.g.][]{robutel,marzari13,dvorak07, freist, erdi13} and the capture of 
Trojan asteroids \cite{sch12}. 
\cite{robutel} also investigated the secondary and secular resonances of
 Jupiter's Trojans by means of frequency map analysis.
Some of these studies were dedicated to high inclined Trojan orbits 
(e.g. for Jupiter Trojans \cite{sch04} and \cite{dvorak05}, for Neptune Trojans \cite{zhou} 
and for Uranus Trojans \cite{dvorak10}). \cite{erdi84} studied the critical inclination $i_c$ 
of Trojan asteroids analytically.
In a more general work of \cite{brasser} they derived a 
formula\footnote{The dependence on mass is rather weak for planetary size 
bodies} for the critical inclination as a function of the mass of the secondary 
body (planet):
\begin{equation}
\mathrm{i_{c}}\approx {61.5^{\circ}(1-1.123\mu)}.
\end{equation}

Later investigations on high inclined orbits like by \cite{funk12} could confirm 
the results on $i_c$. In addition they found stable orbits (of the Trojans and
the gas giant) for eccentricities up to 0.6 and for two different 
mass ratios $\mu=0.001$ and $\mu=0.007$.
 Close to these stability limits the Kozai resonance appeared, which is shown in 
the work of \cite{zhou} and \cite{funk12}.

An interesting topic is also the possibility of Trojan planets in extrasolar 
planetary systems like it is discussed e.g. in the dynamical investigations of 
\cite{nauen}, \cite{erdi05}, \cite{dvorak04}, and \cite{sch07}.

In our Solar system, the eccentricities and inclinations of Trojan asteroids 
range widely and are often far greater than that of their host planet. 
This is discussed in the work of \cite{sch13}, where the stability was
investigated for Trojan asteroids. 
 
In continuation of our former work \citep{sch13} we concentrated our stability 
investigations now on lower mass ratios $\mu \le 0.001$. Therefore we investigated 
the stability of the Lagrangian point $L_4$ in the spatial restricted three-body 
problem to determine the stability limits for different masses of the 
secondary body. In addition we considered secondary resonances and their influence 
on the stability.

\section{Models and methods}
\label{setup}
We studied the spatial restricted three-body problem (SR3BP). In this problem 
two finite bodies, the primaries, revolve about their common centre of mass, 
and a third massless body moves under their gravitational influence not confined 
to the orbital plane of the primaries. We used the SR3BP in its variants: 
circular or elliptic spatial restricted three-body problem (C-SR3BP or E-SR3BP) 
depending on the orbit of the primaries. To study the stability in the equilibrium 
point $L_4$, for both cases -- in the C-SR3BP and E-SR3BP -- we integrated the 
equations of motion for the stability maps up to $T_c=10^6$ periods of the primaries 
by using a Bulirsch-Stoer integrator and the Lie integration method complementary.
For the boundaries between stable and chaotic regions or sticky orbits 
we integrated up to $T_c=10^8$ periods.
We also made long term calculations for a sample of resonant and non resonant 
orbits.
 
For the stability analysis we used the method of the maximum eccentricity 
$e_{max}$ and the Lyapunov characteristic indicators (LCI). The results of the 
LCI and $e_{max}$ are in a good agreement as shown in \cite{sch07,sch09}. The LCI is 
the finite time approximation of the maximal Lyapunov exponent.
For the maximum eccentricity $e_{max}$ we checked the largest eccentricity of
the test particle during its motion. For larger eccentricities it is more 
probable that the orbit of the test particle becomes chaotic (having close
encounters or even collisions with the primaries). 

In order to obtain stability maps in the circular case, depending on the 
mass parameter $\mu$ of the primaries\footnote{We decided to use the common 
definition of the dimensionless mass ratio $\mu$ as given in the introduction, and the 
dimensional mass parameter given in Earth-masses to compare more easily with the solar system} 
and the orbital inclination $i$ of the test particle, we changed the mass parameter 
$\mu$ between $1 \leq m_{Earth} \leq 1400$ (1400 Earth-masses is 
equal 4.5 Jupiter-masses) with a stepsize $\Delta m_{Earth}=2$, and the initial inclination 
$i$ between $0^{\circ} \leq i \leq 60^{\circ}$ with a stepsize $\Delta i = 10^{\circ}$. 
In the elliptic case (E-R3BP), we changed $\mu$ in the same way,  $i$
between $0^{\circ}\leq i \leq 60^{\circ}$ with a stepsize $\Delta i = 10^{\circ}$, 
and the orbital eccentricity $e$ of the primaries between $0 \leq e \leq 0.99$ with a 
stepsize $\Delta e=0.01$. Please note that the initial eccentricity of the test particle 
was always set to that of the secondary and we set $\varpi=60^{\circ}$ and $M=0$. 

As a consequence the test particle always started in $L_4$.  
The stability maps show either $e_{max}$ or LCI for each orbit corresponding to the 
above mentioned initial conditions.

The classification of the orbit was done by checking the libration 
amplitude $\sigma$ which is defined as the difference between the mean 
longitude of the asteroid and the planet ($\lambda_{Tro} -\lambda_P$). 
$\lambda_{Tro}$, $\lambda_P$ are given by $\lambda_{Tro}=\varpi+M$, 
$\lambda_P=\varpi_P+M_P$ where $\varpi$, $\varpi_P$ are the longitudes of 
pericenter of the massless body and of the planet and $M_{Tro}$, $M_P$ 
are the mean anomaly of the third body respectively of the planet. This
was already used in the work of~\cite{sch13}. The difference to our former study is
that we fix the initial eccentricity of the Trojan to that of the secondary body.

Another important point in this work is the analysis of the frequencies of the 
librational motion to determine secondary resonances. In a more general study of 
resonances~\cite{robutel} did this for the Jupiter-Saturn system (fixed mass ratio).
We computed also the frequencies of librational motions around $L_4$ and determined 
secondary resonances. 
In the planar, circular restricted three-body problem there are two frequencies, 
$n_s$ and $n_l$, corresponding to the short and long period components of libration. 
In the planar, elliptic restricted three-body problem $L_4$ moves in an elliptic 
orbit itself and its mean motion $n$  combines with $n_s$ and $n_l$. Therefore, in 
the elliptic case there are four frequencies of libration
$n_s$, $n_l$, $n-n_l$, and $n-n_s$. \cite{rabe1,rabe2} gave the normalized 
frequencies $n_s$ and $n_l$ taking $n$ as unit frequency. \cite{erdi07,erdi09} 
studied secondary resonances between the four frequencies and introduced several 
types. Type A corresponds to $(1-n_l):n_l$, type B to $n_s:n_l$.

To compute the frequencies of libration in the spatial case we used the 
Laplace-Lagrange variables $h,k,p,q$ and the angle of libration $\sigma$ given by
\begin{itemize}
\item $h =e \sin (\omega + \Omega)$,  $k= e \cos (\omega + \Omega)$ \\
\item $p= \sin i \sin \Omega$,  $q= \sin i \cos \Omega$, \\
\item $\sigma= \cos (\lambda - \lambda')$,  $\lambda = \omega + 
\Omega + M$.\\
\end{itemize}
Here $e$ is the eccentricity, $i$ the inclination, $\omega$ the argument of 
pericentre, $\Omega$ the ascending node, $\lambda$ the mean orbital longitude of 
the orbit of the test particle, and $\lambda'$ is the mean orbital longitude of the 
smaller primary. 
(The angular orbital elements refer to the orbital plane of the primaries as 
reference plane, in which the reference direction points to the pericentre of the 
relative orbit of the primaries.) 

Placing the test particle in $L_4$ with initial parameters described earlier, we 
followed its orbit for $10^6$ periods of the primaries by using the Lie integrator 
and computed the corresponding $h,k,p,q$ and  $\sigma$ variables to which we applied 
a discrete Fourier transform (DFT, with SigSpec), or a fast Fourier transform 
(FFT, via FFTW) to find the peaks in the power spectrum. 
The locations of the maxima are interpolated via a quadratic 
function, taking into account the three highest samples around the tentative 
peak. 
SigSpec computes the spectral significance levels for the DFT amplitude 
spectrum of a time series at arbitrarily given sampling. It solves for the 
probability density function of an amplitude level, including dependencies on 
frequency and phase \citep{reegen}.
Frequency map analysis was used to investigate the stability properties of tadpole 
orbits in the proximity of the equilibrium points of Jupiter in the work of~\cite{marzari03}
and \cite{robutel}.

\section{Stability}
\label{stb}

\begin{figure*}
\includegraphics[width=6.1cm,angle=0]{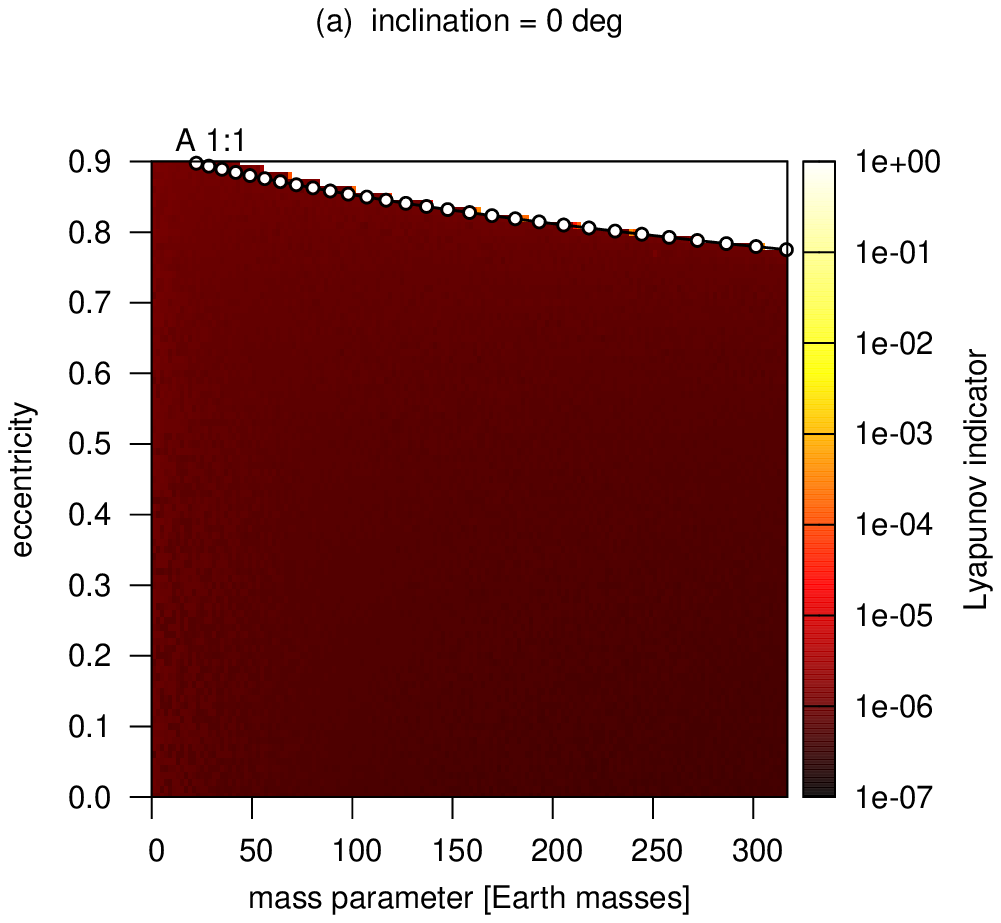}
\includegraphics[width=6.1cm,angle=0]{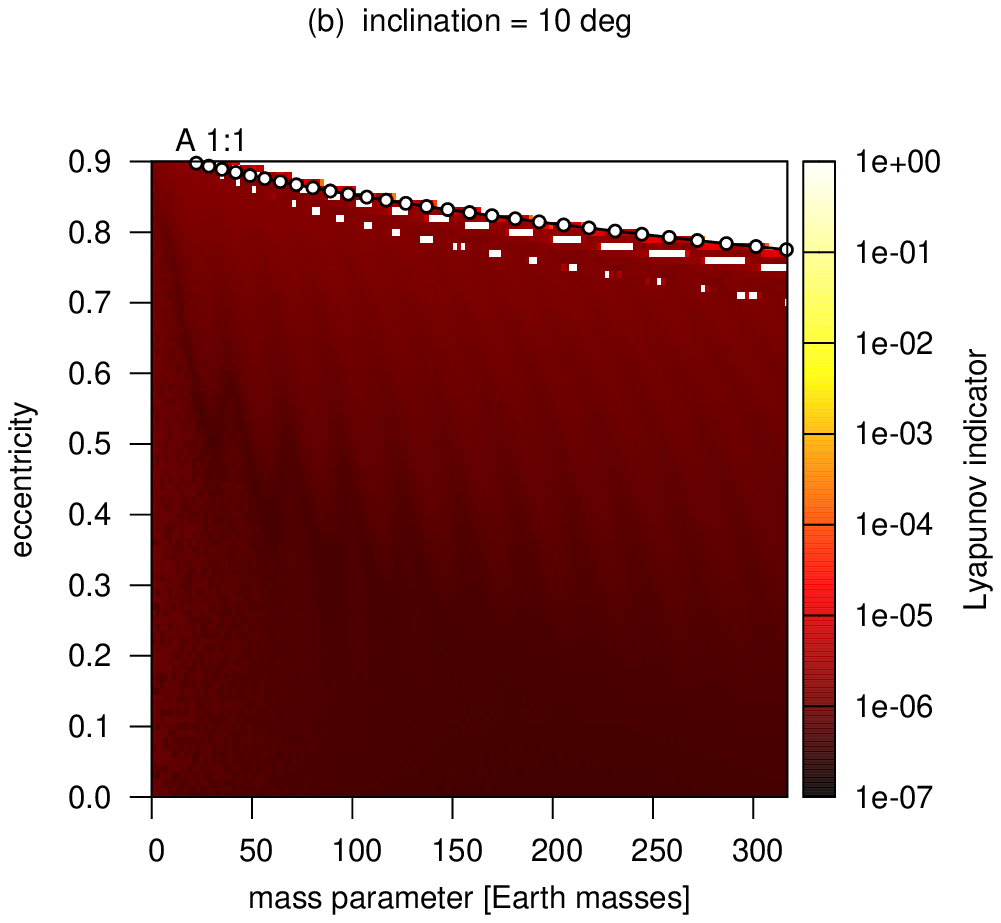}
\includegraphics[width=6.1cm,angle=0]{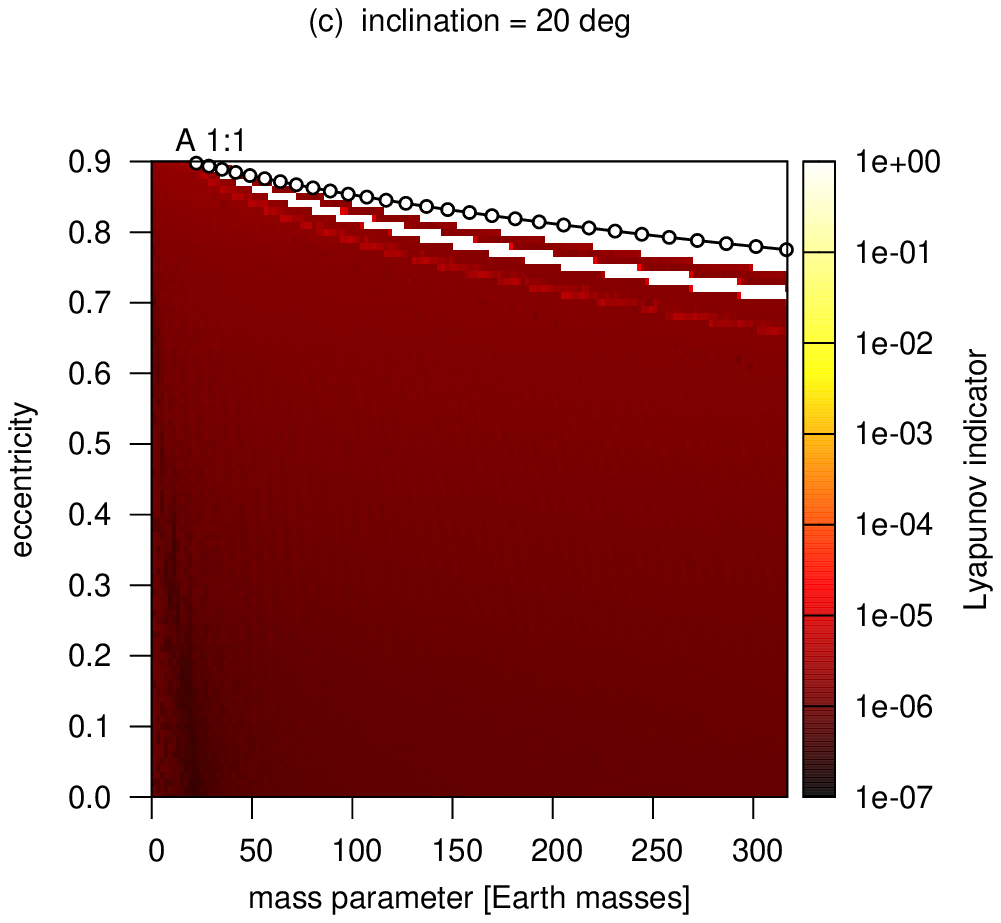}
\includegraphics[width=6.1cm,angle=0]{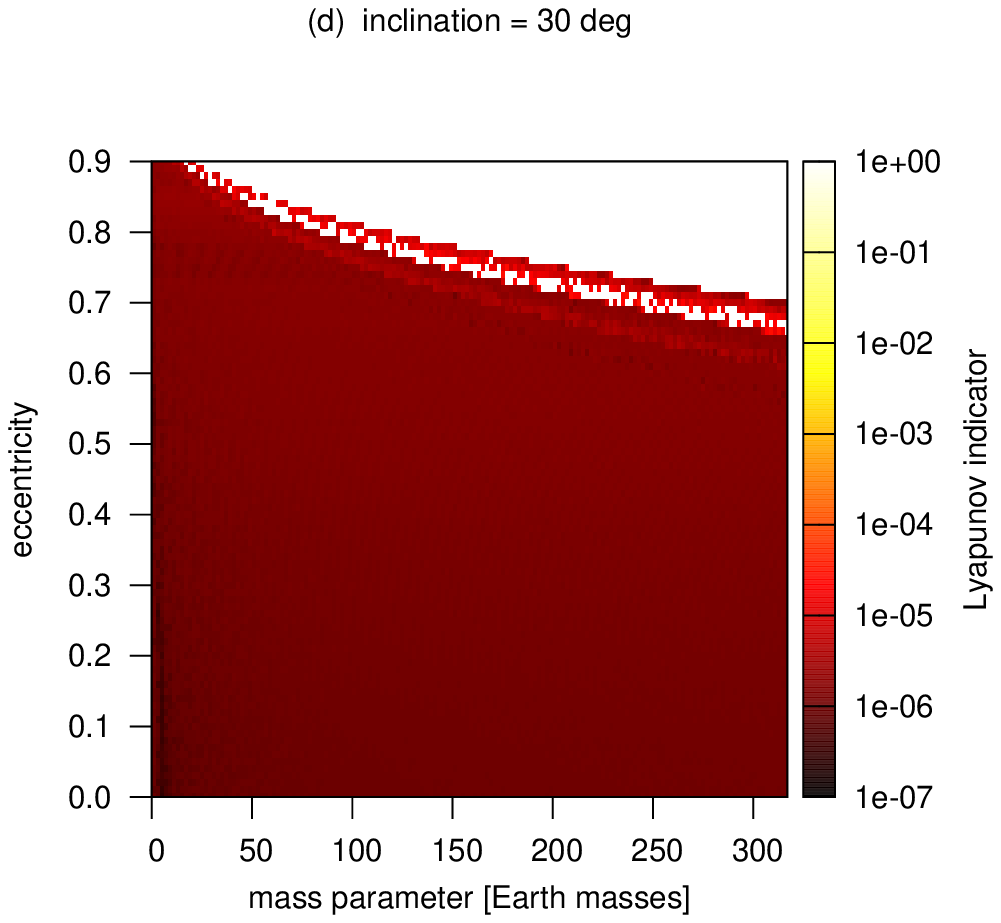}
\includegraphics[width=6.1cm,angle=0]{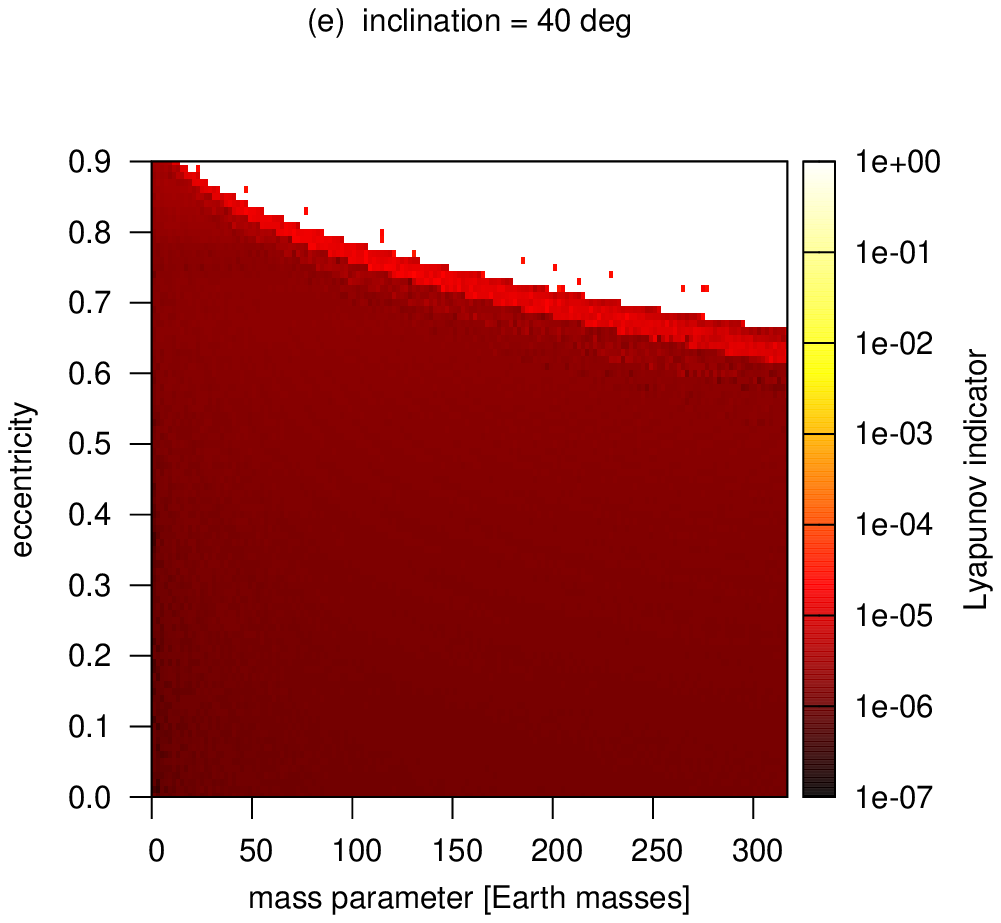}
\includegraphics[width=6.1cm,angle=0]{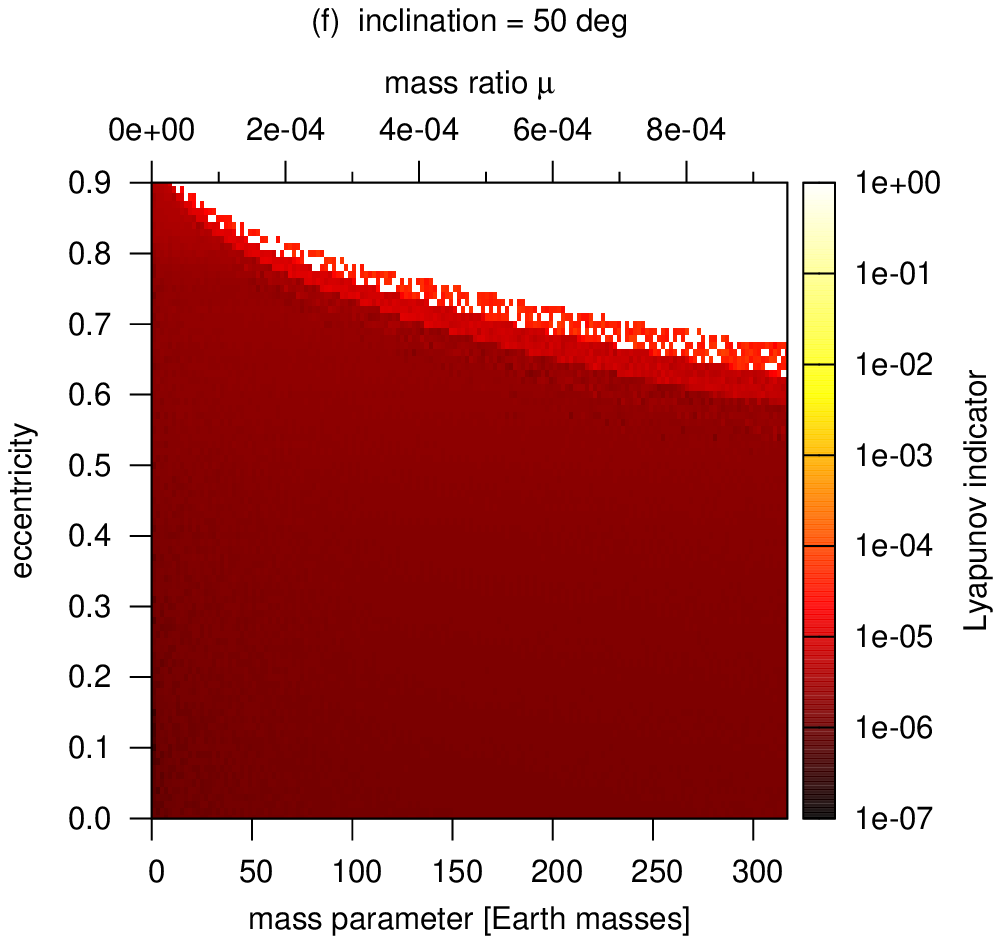}
\caption{The LCI maps showing the stability of  $L_4$ in the E-SR3BP
depending on the eccentricity $e$ and the mass parameter $\mu$ given in 
Earth masses, and for comparisons as mass ratio as defined in chapter\ref{intro}
(lower right graph). The dark red color depicts stable motion, whereas the yellow
and white colors represent the chaotic behaviour. For color plots see online version.
The circles in black and white represent the analytical curve of the A~1:1 
resonance (given in the upper graphs and the middle left one).}
\label{fig1}
\end{figure*}

In this chapter we concentrate our studies on the stability of
the Lagrangian point $L_4$ in the spatial restricted three-body problem, for 
moderate mass ratios $\mu$, meaning that $\mu \le 0.0045$, which represents the mass 
parameter of up to 1400~Earth-masses ($M_E$) equal to 4.5 Jupiter-masses in the Solar system.
The LCI stability maps are given in Fig.~\ref{fig1} for different inclinations 
$(i=0, 10, 20, 30, 40, 50^{\circ})$, starting with the planar case (Fig.~\ref{fig1} 
upper left graph). The graphs of Fig.~\ref{fig1} for higher inclinations show a structure
that shift to smaller eccentricities. We could allocate the structure as the A~1:1 resonance.
For comparison we included the well known analytical curve of the A~1:1 resonance \citep{erdi07} 
for the initial inclinations $i=0, 10, 20^{\circ}$ 
(Fig.~\ref{fig1} upper graphs and middle left graph) of the Trojan bodies.
The curve (circles in black and white) fits well for $i=0, 10^{\circ}$, but does 
not fit anymore for $i=20^{\circ}$ and higher inclinations. This is because of the 
larger chaotic regions, which can be seen in Fig.~\ref{fig1}, middle and lower graphs 
($i=20-50^{\circ}$); $i=60^{\circ}$ is not presented, because there are
only a few stable orbits. 

For the conclusion we counted the number $n$ (in percent) of stable orbits. 

\begin{table}
  \caption{The percentage of the number $n$ of stable orbits for each stability map derived from 
Fig.\ref{fig1}.}
  \begin{tabular}{lll}
  \hline
&inclination [$^{\circ}$] & $n$ of stable orbits [\%] \\
\hline
&	0	& 93 \\
&	10	& 92 \\
&	20	& 89 \\
&	30	& 86 \\
&	40	& 84 \\
&	50	& 83 \\
&       60      & 19 \\
\hline
\end{tabular}
\label{tab1}
\end{table}

The results are presented in Tab.~\ref{tab1} for the different stability maps for
different initial inclinations ($i=0^{\circ}-60^{\circ}$).
We can conclude that the number of stable orbits for inclinations between 
$i=0^{\circ}$ and $i=50^{\circ}$ is almost constant and shrinks rapidly for $i=60^{\circ}$. 
The number of stable orbits can be given as the function of the
 inclination by a semi-empirical formula: $n=-0.22i+93$.
In addition we studied the variation of $e_{max}$. The stability maps gave the same results, 
therefore they are not shown here. 

In addition we analysed the libration amplitude, because in the case of a large libration 
amplitude the Trojan may get on a horseshoe orbit or may be ejected out of the Trojan 
region after a close approach with the planet (secondary body). The value of $\sigma$
should be smaller than $180^{\circ}$ for tadpole orbits. Former studies \citep[see][]{sch13} 
showed -- by the help of the libration amplitude $\sigma$ -- that not all 
stable asteroids are moving in Trojan motion (= tadpole orbits).
Therefore we checked the planar case and made a cut of our stability map shown Fig.~\ref{fig1} 
(upper left graph) in 0.001 (1 Jupiter mass) and 0.000117 (40 Earth masses) for $10^8$ years.
In addition we checked the libration amplitude for different inclinations for $\mu=0.001$ shown in 
Tab.~\ref{tab1b}. We verified that the critical angles 
-- we also call it libration amplitude $\sigma$ -- are oscillating around 
$\lambda_{Tro}-\lambda_{P}=60^{\circ}$ with a very small amplitude $\sigma=10^{-6}$, even for
e=0.75 and $\mu=0.001$ (shown in section~\ref{appendix}).

 \begin{table}
  \caption{The maximum eccentricity of test objects moving in stable tadpole orbits 
($\sigma << 180^{\circ}$).}
  \begin{tabular}{lll}
  \hline
&inclination [$^{\circ}$] & maximum \\
& &           eccentricity\\
\hline
&	0	& 0.77 \\
&	10	& 0.73 \\
&	20	& 0.69 \\
&	30	& 0.64 \\
&	40	& 0.62 \\
&	50	& 0.61 \\
&       60      & 0.02 \\
\hline
\end{tabular}
\label{tab1b}
\end{table}

\section{Secondary resonances}
\label{res}
Complementary to the stability analysis we made a frequency analysis to
find new secondary resonances.
Again we studied the stability maps depending on the orbital eccentricity $e$ 
of the primaries and the mass ratio $\mu$ of the primaries. 
We placed the test particle in $L_4$ and started it on inclined orbits. 
Changing $e$ and $\mu$ as described in Section~\ref{setup}, we followed the orbits 
for only $10$ periods of the primaries and computed stability maps in the $\mu$-$e$ 
plane by recording the $e_{max}$ and LCI values for each $(\mu,e)$ pair. 
 
\begin{figure*}
\includegraphics[width=10.0cm,angle=0]{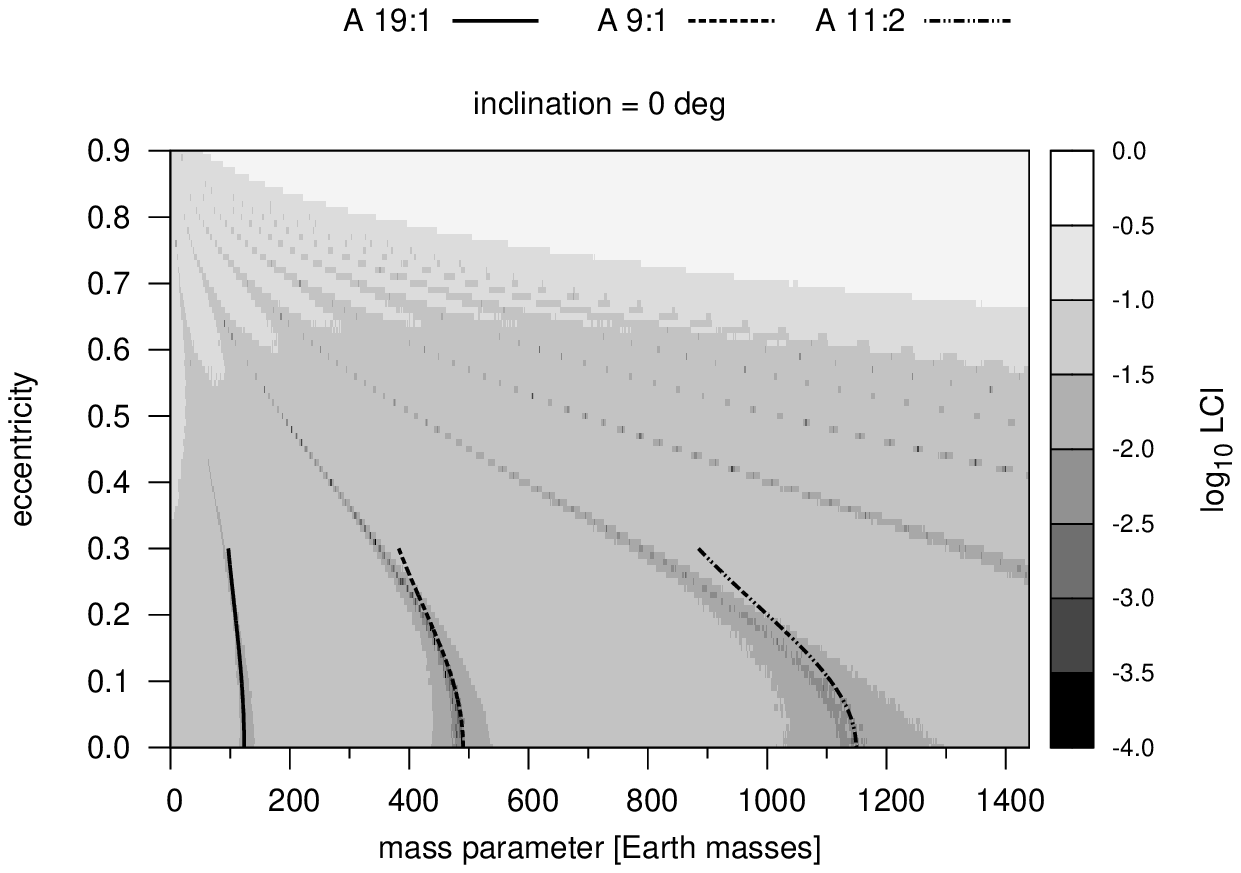}
\includegraphics[width=10.0cm,angle=0]{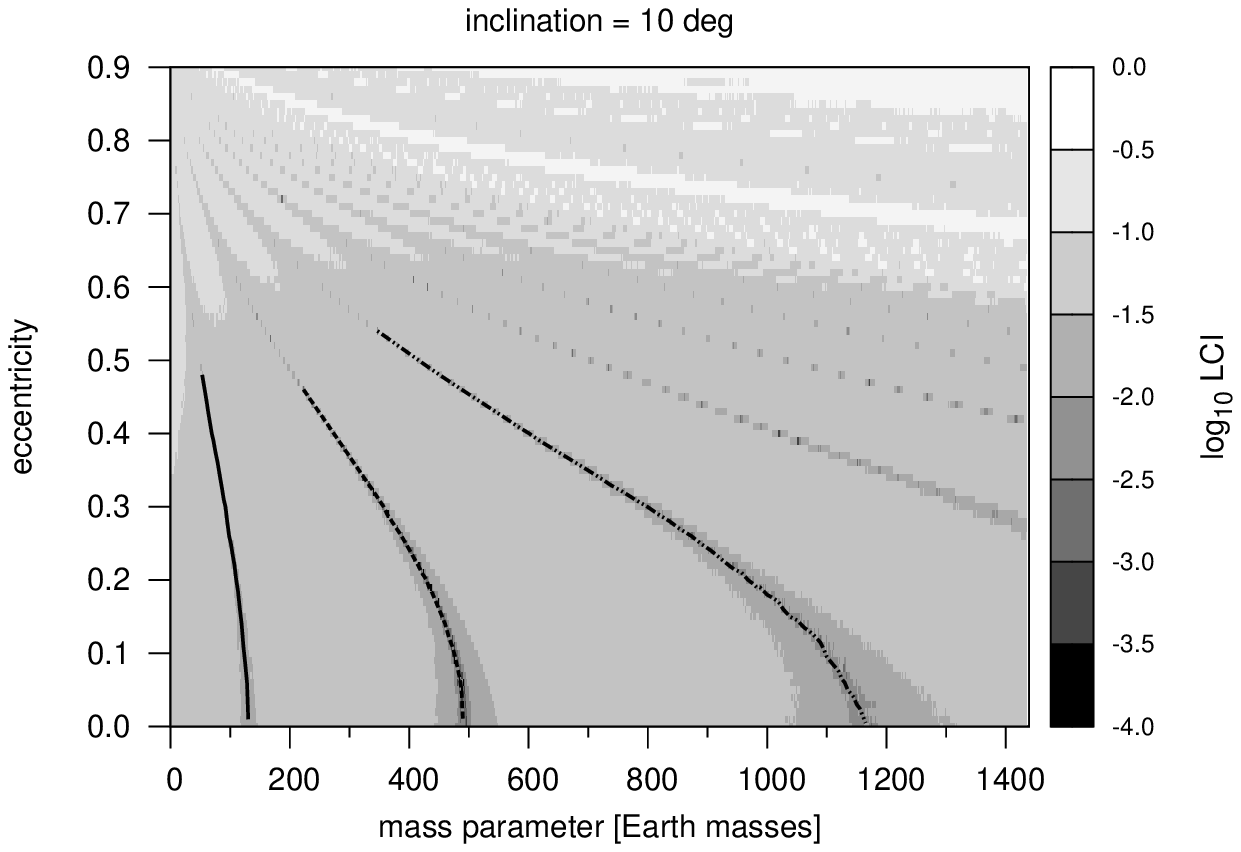}
\includegraphics[width=10.0cm,angle=0]{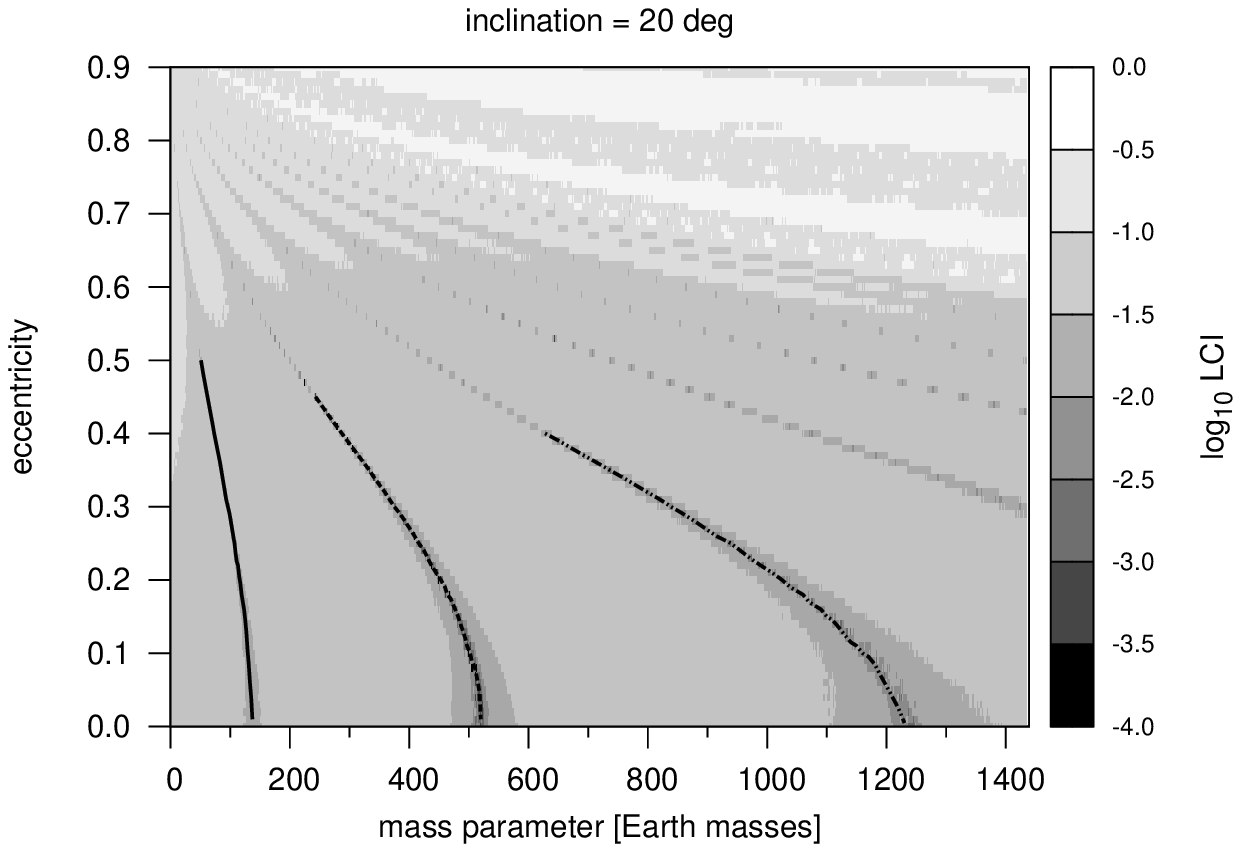}
\caption{The LCI maps showing the stability of $L_4$ depending on the 
eccentricity $e$ and the mass parameter $\mu$. 
The dark grey region depicts stable motion, the light regions correspond to chaotic 
behaviour. The lines represents three different resonances; in case of 
$i=0^{\circ}$ (upper graph) the analytical curves were obtained from Rabe's equation (1970), 
whereas in case of $i=10^{\circ}$ (middle graph) and $i=20^{\circ}$ (lower graph) the curves
presents the results of the frequency analysis.}  
\label{fig2}
\end{figure*}

\begin{figure*}
\includegraphics[width=10.0cm,angle=0]{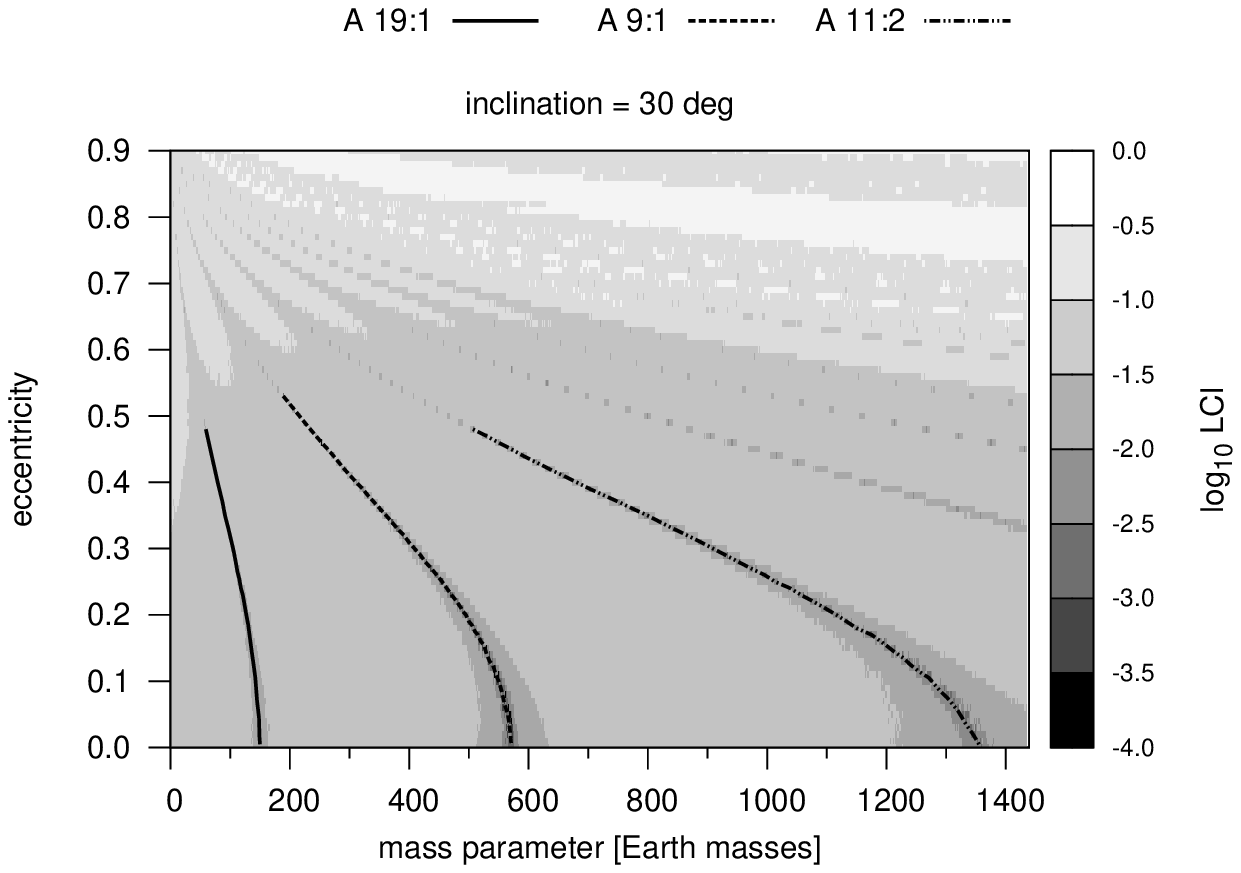}
\includegraphics[width=10.0cm,angle=0]{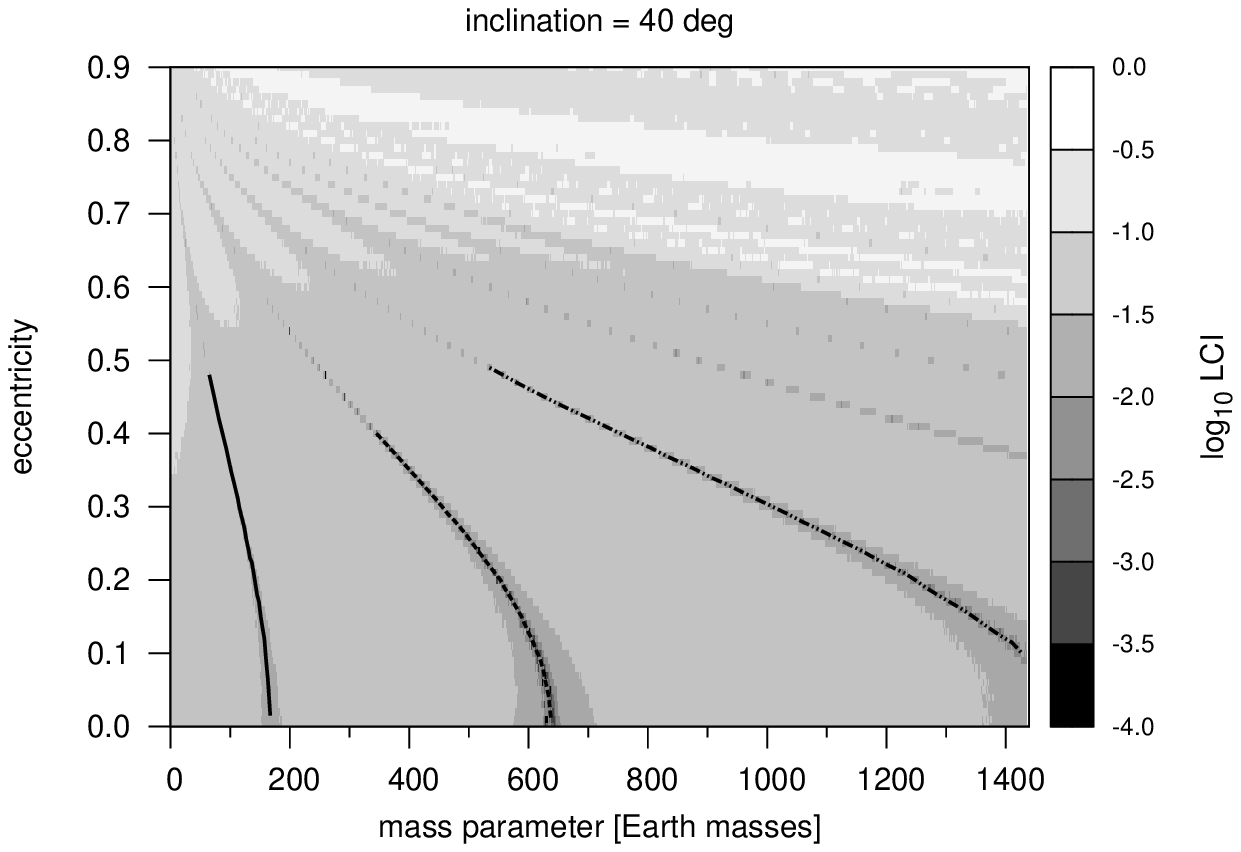}
\includegraphics[width=10.0cm,angle=0]{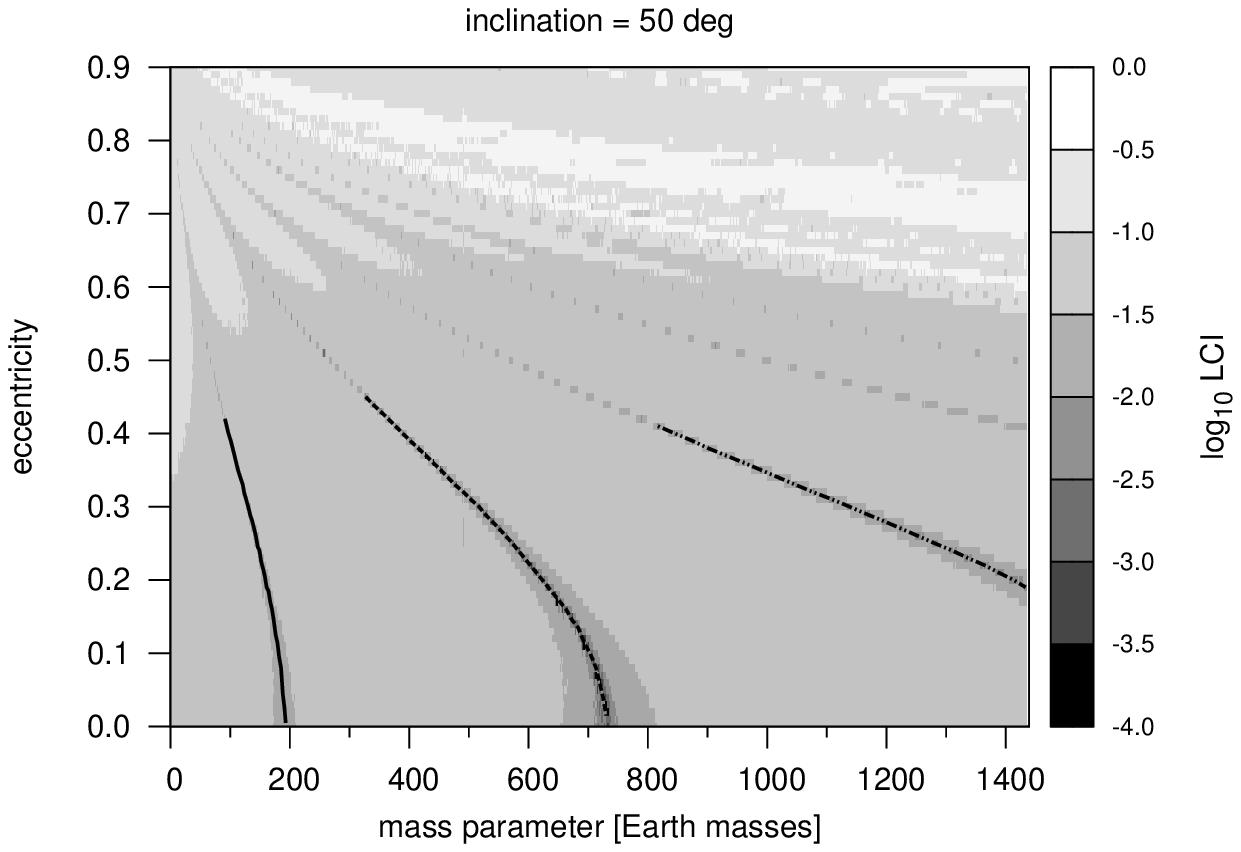}
\caption{The LCI maps showing the stability of $L_4$ depending on the eccentricity 
$e$ and the mass parameter $\mu$. The dark region depicts stable motion, the light regions 
correspond to chaotic behaviour. 
The curves represents three different resonances in case of $i=30^{\circ}$
(upper graph), $i=40^{\circ}$ (middle graph) and $i=50^{\circ}$ (lower graph) obtained by frequency analysis.}
\label{fig3}
\end{figure*}

We studied the stability maps for a very short time ($T_c$=10 periods), because the 
structure of the secondary resonances are blurred for longer integration time. 
The stability maps are presented in Fig.~\ref{fig2} for $i=0, 10, 20^{\circ}$ (upper, 
middle and lower graph) and Fig.~\ref{fig3} present $i=30, 40, 50^{\circ}$, where
the curves (lines) represent three different secondary resonances. 

\begin{table}
  \caption{Initial points of the resonant curves obtained from Rabes equation \citep{rabe2} 
 for the planar circular case (Fig~\ref{fig2} upper left graph).}
  \begin{tabular}{lllll}
  \hline
Type & Resonance & Frequency &	$\mu$ & comment\\
\hline
 A & 19:1&	0.95000000	& 0.00036972 & new\\
 B & 20:1&      0.99875234	& 0.00036880 & ?\\
 A & 9:1&	0.90000000	& 0.00147099 & new\\
 B & 10:1&	0.99503719	& 0.00145652 & ?\\
 A & 11:2&	0.84615385	& 0.00344712 & new\\
 A & 4:1&	0.80000000	& 0.00575455 &\\
 A & 3:1&	0.75000000	& 0.00883461 &\\
 A & 1:1&	0.50000000	& 0.02943725 & Fig.~\ref{fig1}\\
\hline
\end{tabular}
\label{tab2}
\end{table}

In case of $i=0^{\circ}$ (upper graph of Fig.~\ref{fig2}) we show the analytical curves obtained 
from Rabe's equation~\citep{rabe1,rabe2}, and in case of higher inclination we present 
the numerical results (by using frequency analysis). Obviously the 
curves obtained from Rabe's equation are not valid for higher eccentricities, but they are also 
valid for moderate eccentricities as was shown in \cite{erdi07}.
We could also find a shift of the resonances when we studied the spatial case
for large $\mu$ \citep{sch12a}.
Were in Fig.~\ref{fig2} and Fig.~\ref{fig3} the results are shown for small $\mu$.
Before we go into the details we have to mention that the secondary resonances for
small mass parameters are very close together and therefore not easy to locate.
However, we found three new resonances -- the A~19:1, A~9:1 and A 11:2 -- which are shown 
for the planar case the upper left graph of Fig.~\ref{fig2} 
and the initial points are presented in Tab.~\ref{tab2}. We note that the B 10:1 and the B 20:1
are also very close to the curve of A~19:1, but at the current resolution $\Delta \mu=2M_E$ they
are not distinguishable, because of the minimum distance of the resonances which is $\approx 0.5 M_E$. 
However, the structures in the LCI seem to be associated preferably with the A-type resonances. 
These structures are visible in Figs.~\ref{fig2},\ref{fig3} but not labeled, because they are very close.

In addition we could allocate the A~4:1 (curve has an initial point at 
$e \approx 0.3$) and the A~3:1 resonances (curve has an initial point at $e \approx 0.4$). 
The last two resonances are not labeled in Figs.~\ref{fig2},\ref{fig3} and the initial points lie 
outside the scale of the graphs, because of their large mass ratio as shown in 
Tab.~\ref{tab2}. 

As mentioned before, the secondary resonances shift to higher mass parameter when the 
inclination increases. This can be seen clearly when we look at the initial points of the
resonances ($e=0$). The resonance A~19:1 for the planar case at 160~$M_E$ is shifted to
200~$M_E$ ($i=50^{\circ}$) see Fig.~\ref{fig2}, which is a difference of $\Delta \mu=40 M_E$.
The initial resonant point of the resonant curve A~9:1 is shifted from 490~$M_E$ to
720~$M_E$ ($\Delta \mu=230 M_E$) and for A~11:2 we measure  a shift from 1150~$M_E$ to 
1700~$M_E$ (lower graph Fig.~\ref{fig3}) with a difference $\Delta \mu=550 M_E$.
However, the shift with larger inclinations depend on the starting position of the 
resonance. That means that for low mass parameters the shift is smaller (A~19:1, 
$\Delta \mu=40 M_E$), whereas for larger values of $\mu$ the shift is larger (A~11:2,   
 $\Delta \mu=550 M_E$).

\begin{figure*}
\includegraphics[width=9.0cm,angle=0]{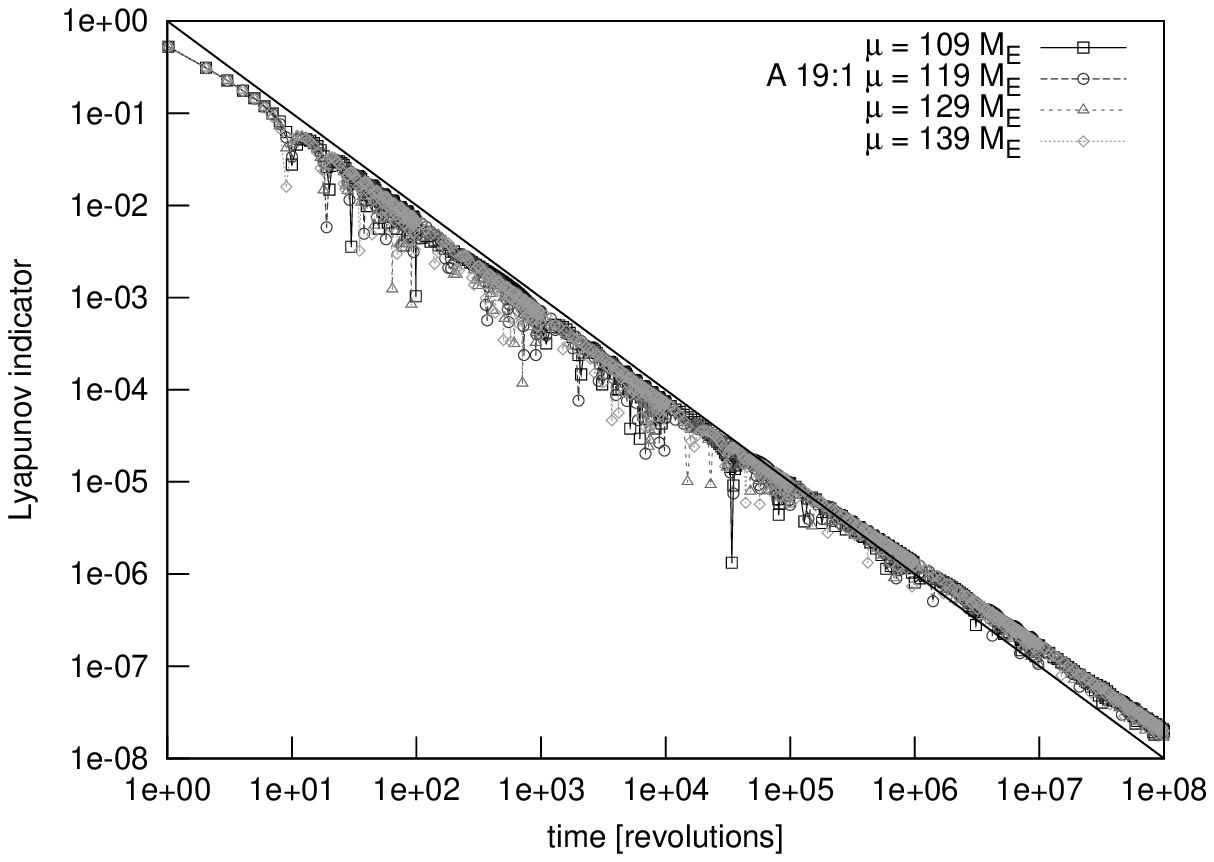}
\caption{The LCI for $T_c=10^8$ periods for 4 different orbits (different masses) in, 
close and out of the secondary resonance A~19:1. The diagonal line represents the theoretical
behaviour of the Lyapunov exponent decreasing inversely proportional with time.}
\label{fig4}
\end{figure*}

Finally, we studied the long term stability of the resonances. For this we integrated
orbits in, close to, and outside the secondary resonances for a moderate initial 
eccentricity ($e=0.2$, shown in Fig.~\ref{fig4}) and a larger value of $e=0.6$. An example is presented in 
Fig.~\ref{fig4}, where we plotted the curve of the LCI for the complete integration time $T_c=10^8$ periods,
for 4 different orbits around the A~19:1 resonance.
We can summarize that there is no difference in and outside the resonance and
have no influence on the stability, but we have to remark that these are high order
resonances. However, this result show that the investigated secondary resonances will 
not play a role in the Solar system. 

\section{Conclusions}
\label{conclusion}

In this paper we studied the stability of the Lagrangian point $L_4$ in the spatial 
restricted three-body problem (SR3BP). 
It is known that in the planar, circular restricted three-body problem $L_4$ is 
linearly stable for mass ratios $\mu \lesssim 0.0385$. However, we concentrated our 
investigations on small mass ratios $\mu \le 0.001$, which represents the mass 
ratios for stable configurations of tadpole orbits in the Solar system.
We computed different stability maps in the SR3BP by changing the orbital inclination 
($\Delta i=10^{\circ}$) of the test particles. However, the stability maps were computed 
by changing the mass parameter and the ecccentricity of the primaries; for a very fine 
grid of initial conditions ($\Delta \mu=2M_E$ and $\Delta e=0.01$).

The stability analysis had shown that the size of the stable region of the stability
maps ($\mu$-$e$ plane) shrinks with the increase of the inclination (Fig.\ref{fig1}). 
The decrease of the number of stable orbits (in percent of all orbits) 
is almost linear with respect to the increase of the inclination and dropping down 
abruptly at $i=50^{\circ}$; shown in Tab.\ref{tab1}. 
We could confirm -- by computing the libration amplitude -- our former results that 
the stability maps mainly show tadpole orbits. The small libration amplitude follows 
from the initial conditions starting in the equilibrium point $L_4$ with equal eccentricity
of the secondary and Trojan body.

Our investigations showed that there exist also secondary resonances for small $\mu$.
Using Rabe's equation \citep{rabe2} and frequency analysis we could determine
the A-type~19:1, A~9:1 and the A~11:2, which are high order resonances. We found also 
B-type resonances, but at the current resolution they are not distinguishable from the A-type resonances.
We can conclude that these resonances have no influence on the orbital stability (Fig.~\ref{fig4}).
Nevertheless we could show that the secondary resonance A 1:1 is relevant for the stability and 
shifts with larger inclinations.   
We could also show that a few resonances whose initial points lie in the region of higher 
mass ratio (presented in \cite{erdi07}) move into the region of low mass ratios.   
Also the well known analytical curve of the A~1:1 resonance move into the region
of low mass parameter. For initial inclinations $i=0^{\circ}, 10^{\circ}$ 
(Fig.~\ref{fig1} upper graphs and middle left graph) the curve fits well with the 
limit between stable and unstable region, but it does not fit anymore for 
$i=20^{\circ}$ and higher inclinations (the different types of resonances
were defined in \cite{erdi07} for the planar case), because the A~1:1 resonance shifts to 
lower eccentricities.

Further investigations have to be done to investigate the long term stability 
of the secondary resonances and to find new ones.

\appendix
\section{Critical argument}
\label{appendix}

\begin{figure*}
\includegraphics[width=10.0cm,angle=0]{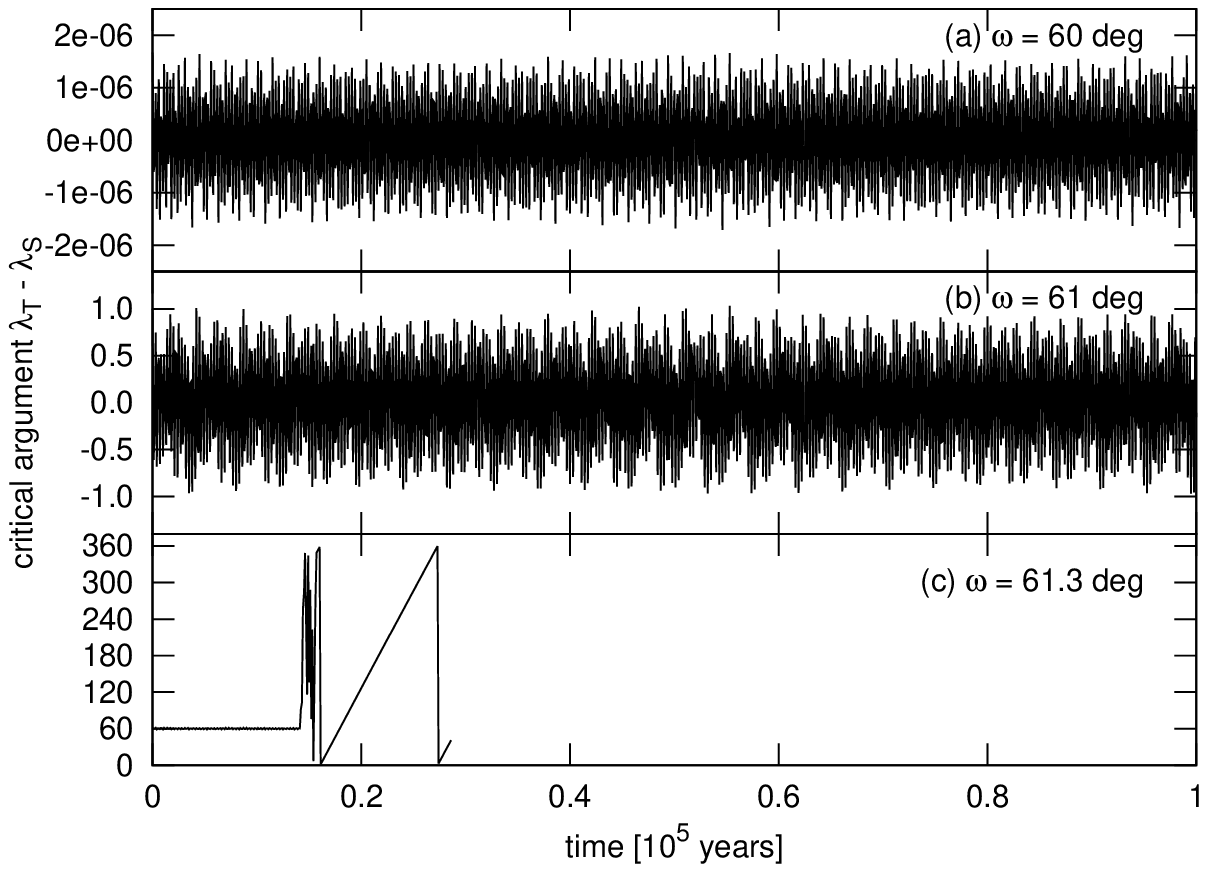}
\includegraphics[width=10.0cm,angle=0]{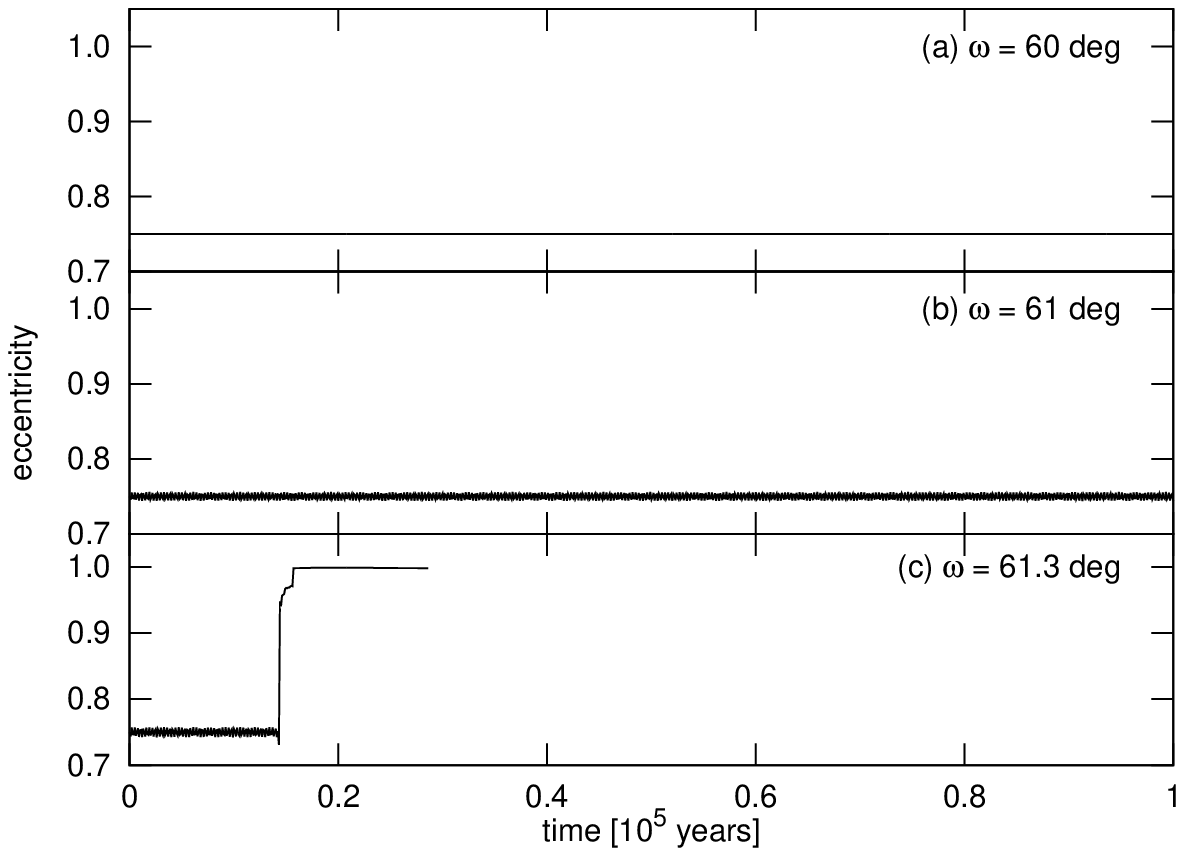}
\caption{The critical argument for libration for a time span of $10^5$ years
  and eccentricity $e$ for the mass parameter $\mu=0.001$. The three cases
  (a-c) show the different argument of perihelion $\omega$, except that the
   initial $e$ of the secondary and the Trojan body is the same.}  
\label{figA1}
\end{figure*}

With Fig.~ \ref{figA1} we checked the critical argument (maximum and minimum
value) for a time spane up to $10^5$ years. We found out that the critical
argument shows small oscillations (about $10^{-6}$). For the cases a) and b)
we can see that the eccentricity is very constant, however for case c) where
we move away from the Lagrangian point we observed chaotic behaviour within
20000 years. We have to remark that our investigations represent the case a).

\section*{Acknowledgments}
R. Schwarz wants to acknowledge the support by the Austrian FWF project P23810-N16. 
\'A. Bazs\'o wants to acknowledge the support from the Austrian FWF project P23810-N16 and the doctoral 
school Planetology: From Asteroids to Impacts and B. Funk wants to acknowledge the support by 
the Austrian FWF project P22603-N16 and P23810-N16.

%\appendix

%\section[]{Large gaps}

%(This appendix was not part of the original paper by
%A.V.~Raveendran and is included here just for illustrative
%purposes. The references are not relevant to the text of the
%appendix, they are references from the bibliography used to
%illustrate text before and after citations.)

\label{lastpage}
\end{document}